\documentclass{article}

\usepackage{arxiv}

\usepackage[utf8]{inputenc} 
\usepackage[T1]{fontenc}    
\usepackage{hyperref}       
\usepackage{url}            
\usepackage{booktabs}       
\usepackage{amsfonts}       
\usepackage{nicefrac}       
\usepackage{microtype}      
\usepackage{amsmath}
\usepackage{lipsum}
\usepackage{graphicx}
\usepackage{subcaption}
\usepackage[symbol]{footmisc}
\usepackage{fancyhdr}

\title{Efficient hybrid modeling and sorption model discovery for non-linear advection-diffusion-sorption systems: A systematic scientific machine learning approach}

\author{
  Vinicius Viena  Santana* \\
  LSRE--LCM \textsuperscript{\dag}, ALiCE\textsuperscript{\ddag} \\
  Faculty of Engineering\\
  University of Porto \\ 
  \texttt{up201700649@edu.fe.up.pt}
  \And
  Erbet Costa \\
  Department of Chemical Engineering\\
  Norwegian University of Science and Technology\\
  \And
  Carine de Menezes Rebello \\
  Department of Chemical Engineering\\
  Norwegian University of Science and Technology\\
  \And
  Ana Mafalda Ribeiro \\
  LSRE--LCM \textsuperscript{\dag}, ALiCE\textsuperscript{\ddag}\\
  Faculty of Engineering\\
  University of Porto \\
  \And
  Chris Rackauckas \\
  JuliaLab \\
  Computer Science and Artificial Intelligence Laboratory \\
  Massachusetts Institute of Technology\\
  \And
  Idelfonso B.R. Nogueira* \\
  Department of Chemical Engineering\\
  Norwegian University of Science and Technology\\
  \texttt{idelfonso.b.d.r.nogueira@ntnu.no}
  }

\begin{document}
\maketitle
\footnotetext[1]{To whom correspondence should be addressed}
\footnotetext[2]{LSRE - LCM (Laboratory of Separation and Reaction Engineering -- Laboratory of Catalysis and Materials, Faculty of Engineering, University of Porto, Rua Dr. Roberto Frias, 4200-465 Porto, Portugal)}
\footnotetext[3]{ALiCE (Associate Laboratory in Chemical Engineering, , Faculty of Engineering, University of Porto, Rua Dr. Roberto Frias, 4200-465 Porto, Portugal)}
\begin{abstract}
This study presents a systematic machine learning approach for creating efficient hybrid models and discovering sorption uptake models in non-linear advection-diffusion-sorption systems. It demonstrates an effective method to train these complex systems using gradient-based optimizers, adjoint sensitivity analysis, and JIT-compiled vector Jacobian products, combined with spatial discretization and adaptive integrators. Sparse and symbolic regression were employed to identify missing functions in the artificial neural network. The robustness of the proposed method was tested on an in-silico data set of noisy breakthrough curve observations of fixed-bed adsorption, resulting in a well-fitted hybrid model. The study successfully reconstructed sorption uptake kinetics using sparse and symbolic regression, and accurately predicted breakthrough curves using identified polynomials, highlighting the potential of the proposed framework for discovering sorption kinetic law structures.
\end{abstract}

\keywords{Hybrid Modeling \and Scientific Machine Learning \and Sparse Regression \and Advection-Diffusion-Sorption \and Partial Differential Equations}

\section{Introduction}
Mathematical modeling has been an essential part of science and engineering since its early stages. The use of mathematical models for problem-solving in chemical engineering has become increasingly important due to the exponential growth of computer power. Two main approaches have been used to model systems in chemical engineering: The mechanistic/classical and the empirical approaches \cite{RASMUNSON2014}. The classical approach encompasses conservation equations, transport, and thermodynamic expressions and can accurately describe the problem. However, simplifications are often employed to make the problem computationally tractable. The empirical/data-driven models do not make any assumptions about the physics behind the observations, making them flexible. Nevertheless, they are suitable for use in a limited range of process conditions due to their limited extrapolation power \cite{Sansana2021}. 

\begin{figure}[!ht]
    \centering
    \includegraphics[scale = 0.7]{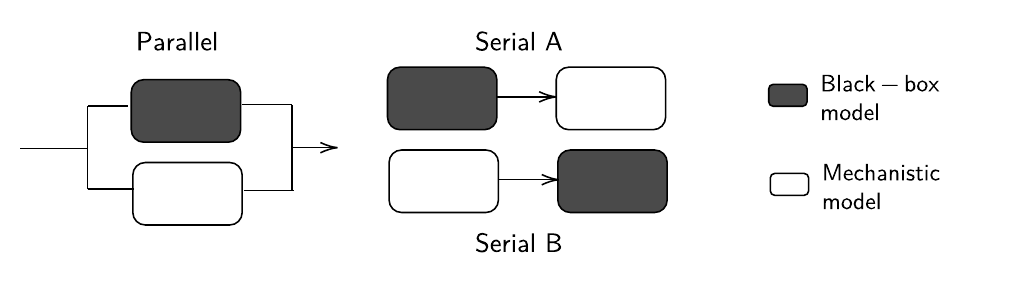}
    \caption{Hybrid models types - adapted from von Stosch et al. (2014) \cite{moritz2014}}
    \label{fig:1}
\end{figure}

On the other hand, empirical models have gained substantial attention in the last few years with machine learning and deep learning. At the same pace, it is becoming more accepted that mechanistic model errors caused by simplifications can be compensated with other forms of knowledge, e.g., empirical models \cite{FEYO1997}. In this sense, hybrid models have received considerable attention in engineering \cite{Laura2020, Pan2022}. The "hybrid" term refers to a structure where empirical equations and physics knowledge (conservation laws, operation invariance, etc.) are merged. The merging process for engineering applications can take several forms, according to von Stosch et al. (2014) \cite{moritz2014}. \autoref{fig:1} illustrates the types of hybrid models according to the authors. The black box represents data-driven models/universal approximations, and the white box mechanistic models are derived from the first principles. The authors divided hybrid models into 3 categories: parallel, serial A and B. Parallel models are suitable when the model structure is not sufficiently known. An empirical model compensates for non-explained prediction residuals. When the model structure is well known, serial A replaces a missing model term with a universal approximator and is expected to perform significantly better than the parallel, especially in extrapolation. Serial B uses mechanistic model predictions as inputs of an empirical model. However, it has found very few applications in chemical engineering. Sansana et al. (2021) \cite{Sansana2021} have extended this classification and include surrogates as a separate class. Another noteworthy class of hybrid model is the approach called Physics-Informed Neural Networks \cite{Raissi2019}. It is conceptually distant from all described hybrid models as it uses physics to regularize Artificial Neural Network (ANNs) training. 

Hybrid serial A models in chemical engineering may combine first-principles derived equations with a universal approximator to replace one or more terms of the equation - kinetics, transport, fluxes. When the underlying physics can be described by differential equations (ordinary, algebraic, or partial), the resulting mathematical object is named Universal Differential Equation (UDE) as described in Rackauckas (2020) \cite{Rackauckas2020}. A universal approximator is a  parameterized object representing  any  possible function in some parameter size limit \cite{Rackauckas2020}. In this sense, Artificial Neural Networks (ANNs) are particularly interesting as they can accommodate high-dimensional input space functions and be made differentiable with appropriate activation functions.

The most common applications of hybrid serial A models lie in bio(reaction) engineering, where the kinetic reaction law is replaced by an ANN \cite{FEYO1997, Zander1999}. However, fitting ANNs' parameters as part of a differential equation poses a significant computational challenge \cite{FEYO1997}. Therefore, applications have been limited to simple systems with few equations and poorly explored for more complicated problems, e.g., distributed-parameter representations such as fixed-bed reactors and chromatographic columns. Recently, Narayanan et al. \cite{Narayanan2021, Narayanan2021B} pioneered the field using hybrid serial A models in nonlinear chromatography where an ANN in a hybrid advection-diffusion-sorption PDE was trained. These works demonstrated the superior capacity of the hybrid model for predicting breakthrough observations and uncovering adsorption equilibrium information. However, despite being able to train the neural networks with Bayesian optimization, it is well known that derivative-free methods such as Bayesian optimization perform poorly in high-dimensional parameter spaces, e.g., artificial neural networks \cite{Moriconi2019,Frazier2018}.

In such scenarios, gradient-based or hessian-based optimizers are more appropriate and preferred. In fact, the state-of-the-art methods for training neural networks are based on gradients \cite{Kingma2015}. In this sense, Nogueira et al. (2022) \cite{Nogueira2022} employed forward-mode automatic differentiation for training an ANN in a hybrid 1D advection-diffusion-sorption PDE. The work demonstrated that it was possible to identify a hybrid model with discrete events that outperformed the purely mechanistic approach. Another pioneering work in adsorption modeling is presented by Praditia et al. (2021) \cite{Praditia2021}, where a FInite volume Neural Network (FINN) is presented for modeling subsurface contaminant transport as a non-linear diffusion-sorption process -- fitting as a hybrid model by capitalizing finite volume structural knowledge and the flexibility of ANNs.

Despite being able to use gradient-based optimizers in Nogueira et al. (2022) \cite{Nogueira2022}, several issues are still open in ANN-based hybrid advection-diffusion-sorption PDE problems. One key problem with using gradient-based optimizers is the ability to calculate the gradient itself accurately and efficiently. The simplest approach is numerical differentiation with finite differences. However, roundoff errors can become prohibitively high and degrade optimizers' performance. The preferred approach is Automatic Differentiation (AD), where high-accuracy gradients can be obtained.
On the other hand, they can be computationally demanding and numerically unstable for non-linear advection-diffusion-sorption PDE problems \cite{Ma2021}. A general guideline for choosing AD for stiff models resulting from PDE discretization and efficiency/stability trade-offs is presented by  Ma et al. (2021) \cite{Ma2021}. Nonetheless, stability and efficiency are system dependent and should be assessed for each problem. Here a guideline for non-linear advection-diffusion-sorption PDE problems is provided - an important contribution for hybrid modeling in sorption processes.

In this sense, the present work demonstrates a feasible and efficient way for training  hybrid non-linear advection-diffusion-sorption PDE problems using gradient-based optimizers via continuous adjoint sensitivity analysis using quadrature adjoint with JIT-compiled tape-based vector jacobian product, orthogonal collocation on finite element PDE discretization scheme, and a fixed-leading coefficient adaptive-order adaptive-time BDF method ODE integrator. As far as we are concerned, it is the first work assessing the numerical aspects of hybrid non-linear advection-diffusion-sorption PDE problems. Once the stability of the hybrid model identification approach is demonstrated, another open issue in the literature is finding ways to make ANNs' predictions interpretable, i.e., identifying the typical functional forms that are expected within domains. One way to achieve it is by using symbolic or sparse regression on the ANNs' output \cite{Rackauckas2020}.  This approach reduces computational burden by avoiding solving differential equations during function space search (as in traditional symbolic regression involving differential equations) and allows finding a combination of simple functions that share similar properties with the trained ANN. A number of studies employed either sparse or symbolic regression on chemical engineering \cite{MCKAY1997981, Cozad2018, NARAYANAN2022133032}. Still, as far as we know, none have used them for giving interpretability to trained hybrid non-linear advection-diffusion-sorption PDEs.   

To demonstrate the contributions above, an in-silico data set is used. The data set was built by simulating breakthrough curves of a hypothetical single-component non-linear advection-diffusion-sorption system with two different isotherms (Langmuir and Sips) and three kinetic models -- Linear Driving Force (LDF), Vermeulen’s model \cite{Hai2004STUDYOT}, and improved LDF \cite{Li2018}. The present work can help engineers to efficiently introduce hybrid modelling in packed-bed separation for either improving predictive power or discovering mass-transfer kinetics direct from breakthrough curves data. 

The present work is organized as follows. Section \ref{sec:methods} presents the methodology, including the \textit{in-silico} data set building process, hybrid model proposition, numerical aspects of the hybrid model solution and gradient calculation, and sparse regression details. Section \ref{sec:results} presents the results for the universal differential equation training and test sets performance and sparse regression on the trained ANN's.

\section{Materials and Methods}
\label{sec:methods}
\subsection{Data set}
\label{sec:dataset}
For demonstrating the aforementioned objectives, an \textit{in-silico}  data set was created by simulating column chromatography breakthrough curves (5.5 mg/L concentration). It was considered an additive Gaussian noise parametrized with a standard deviation compatible with real experimental data (5\%) as referred to in the literature \cite{Nogueira2022, Narayanan2021, Narayanan2021B}. Two types of isotherms -- Langmuir and Sips, three types of adsorption kinetic models -- LDF, Vermeulen's \cite{Hai2004STUDYOT}, and improved LDF from \cite{Li2018} and one sampling rate -- $0.5$ min$^{-1}$ were investigated. The objective is to show the robustness of the proposed method to accommodate several kinetic models and isotherms. The reason for using 5.5 mg/L in the feed concentration was to cover a considerable part of the adsorption isotherm in the breakthrough curve considering the case study as detailed below. The model equations and kinetic laws can be found in the \autoref{append}.

 The hybrid model makes no assumptions about the adsorption kinetics but considers that the adsorption isotherm is known and is an input of the ANN. Moreover, all remaining parameters are considered known -- Peclet number and porosity. This approach differs from Nogueira et al. (2022),  Narayanan et al. (2021a), and Narayanan et al. (2021b) \cite{Nogueira2022, Narayanan2021, Narayanan2021B} work where isotherms were not known and fitted with breakthrough data. Here it is tested if the ANN can learn adsorption uptake from single noisy breakthroughs at column outlet and extrapolate in adsorption and desorption well. 

Only the data at the column outlet was used as the training set, i.e., for fitting ANN's parameters. The reason is that data collection at positions aside from column outlets is not a usual practice in the field. Additionally, two test sets were created with simulation to assess the generalization capacity of the model, i.e., whether the model is general enough to extrapolate adsorption and to predict desorption correctly. \autoref{table:1} shows training set configurations and \autoref{table:2} shows the test set. In the test set, inlet concentrations are changed from 5.5 mg/L to 3.58 mg/L to simulate desorption, from 3.58 mg/L to 7.33 mg/L to simulate extrapolated adsorption for Langmuir isotherm, and changed from 5.5 mg/L to 0.75 mg/L to simulate desorption and from 0.75 mg/L to 9.33 mg/L to simulate extrapolated adsorption for Sips isotherm.

\begin{table}[htpb!]
\centering
\caption{Training \textit{in-silico} data set}
\label{table:1}
\begin{tabular}{ccccc}
\hline
\multicolumn{5}{c}{Parameters - $v = 5.1\times10^{-1}$ (dm/min), L = 2.0 (dm), $Pe = 21.0$, $\varepsilon = 0.5$}                                                                                                                                    \\ \hline
Isotherm   & Kinetic      & \begin{tabular}[c]{@{}c@{}}$c_{in}$\\ (mg/L)\end{tabular} & \begin{tabular}[c]{@{}c@{}}t$_{obs}$\\ (min)\end{tabular} & \begin{tabular}[c]{@{}c@{}}f$_{sample}$\\ (min$^{-1}$)\end{tabular} \\ \hline
Langmuir   & LDF          & 5.5                                                       & 110                                                       & 0.5                                                                   \\
Langmuir   & Vermeulen’s    & 5.5                                                       & 110                                                       & 0.5                                                                   \\
Langmuir   & Improved LDF & 5.5                                                       & 110                                                       & 0.5                                                                   \\
Sips & LDF          & 5.5                                                       & 110                                                       & 0.5                                                                   \\
Sips & Vermeulen’s    & 5.5                                                       & 110                                                       & 0.5                                                                   \\
Sips & Improved LDF & 5.5                                                       & 110                                                       & 0.5                                                                                                         
\\ \hline   
\end{tabular}
\end{table}

\begin{table}[htpb!]
\centering
\caption{Test \textit{in-silico} data set}
\label{table:2}
\begin{tabular}{ccccc}
\hline
\multicolumn{5}{c}{Parameters - $v = 5.1\times10^{-1}$ (dm/min), L = 2.0 (dm), $Pe = 21.0$, $\varepsilon = 0.5$}                                                                                                                             \\ \hline
Isotherm   & Kinetic      & \begin{tabular}[c]{@{}c@{}}$c_{in}$\\ (mg/L)\end{tabular}         & \begin{tabular}[c]{@{}c@{}}t$_{obs}$\\ (min)\end{tabular} & \begin{tabular}[c]{@{}c@{}}f$_{sample}$\\ (min$^{-1}$)\end{tabular} \\ \hline
Langmuir   & LDF          & $5.5 \rightarrow 3.58 \rightarrow 7.33$ & 270                                                       & 0.5                                                                   \\
Langmuir   & Vermeulen’s    & $5.5 \rightarrow 3.58 \rightarrow 7.33$ & 270                                                       & 0.5                                                                   \\
Langmuir   & Improved LDF & $5.5 \rightarrow 3.58 \rightarrow 7.33$ & 270                                                       & 0.5                                                                   \\
Sips & LDF          & $5.5 \rightarrow 0.75 \rightarrow 9.33$ & 270                                                       & 0.5                                                                   \\
Sips & Vermeulen’s   & $5.5 \rightarrow 0.75 \rightarrow 9.33$ & 270                                                       & 0.5                                                                   \\
Sips & Improved LDF & $5.5 \rightarrow 0.75 \rightarrow 9.33$ & 270                                                       & 0.5                                                                                                                             \\ \hline
\end{tabular}
\end{table}

The column geometrical and hydrodynamical parameters were chosen based on Li et al. (2018) \cite{Li2018} work. Even though the velocity does affect the mass transfer coefficient for some situations where external mass transfer is dominant, it was not the aim of this work to assess all specific model parts where outputs can vary. The main goal is to show that the proposed approach is a stable, reproducible, and efficient way of fitting hybrid models in non-linear advection-diffusion-sorption PDE problems. Extending the work to accommodate mass transfer flow-rate dependency is straightforward. It only requires including this variable at the input of the ANN and collecting data at several flow rates. All the methodology presented here still applies with such modifications.    

\subsection{Hybrid model}
\label{sub:hybrid}
The proposed hybrid model inherits most of the structure of the traditional "linear driving force" mechanistic fixed bed model in chromatography \cite{Minceva2015}. It encompasses unsteady-state mass transfer between the fluid and solid phases. The model can be derived from a time-dependent 1-dimensional conservation equation in cylindrical geometry. It considers a plug flow, negligible pressure drop, isothermal operation, and uniform radial distribution with no variation in the superficial flow velocity throughout the bed. The hybridization occurs at the material balance that models mass transfer between solid and fluid phases. Thus, it can be written as:

\begin{align}
&\frac{\partial c}{\partial t^*} = -\frac{1-\varepsilon}{\varepsilon}\textrm{ANN}(q, q^*, \theta) - \frac{\partial c}{\partial x^*} + \frac{1}{Pe}\frac{\partial c^2}{\partial x^{*2}}, \\
&\frac{\partial q}{\partial t^*} = \textrm{ANN}(q, q^*,\theta), \\
&\frac{\partial c(x^* = 1, \forall t)}{\partial x^*} = 0, \\
&\frac{\partial c(x^* = 0, \forall t)}{\partial x^*} = Pe(c - c_{inlet}), \\
&c(x^* \in (0,1), t^* = 0) = c_0, \\
&q(x^* \in (0,1), t^* = 0) = q^*(c_0),\\
&q^* = f(c, p),
\end{align}

in which, $t^* = tu/L$ is the dimensionless time, $x^* = x/L$ is the fixed bed dimensionless length, the axial distance from column inlet, $c$ is the liquid phase concentration, $q$ is the solid phase concentration, $q^*(c,p)$ is the solid phase equilibrium concentration (calculated from a known isotherm $f$ with known parameters $p$), ANN is the artificial neural network parametrized by $\theta$, $\varepsilon$ is the bed porosity, $v$ is the interstitial velocity, $Pe$ is the Peclet number. In this work, $q^*(c,p)$, can be either the Langmuir isotherm $q^* = Qkc/(1 + kc)$ or the Sips isotherm $q^* = Qkc^{\alpha}/(1 + kc^{\alpha})$. For all simulations, as shown in \autoref{append}, the parameters of the isotherms were kept constant with parameters $k = 1.80 L/mg$, $Q = 55.54 mg/g$, $\alpha = 1.5$. Langmuir isotherm parameters were based on Li et al. (2018) \cite{Li2018}, and the Sips isotherm was built so that equilibrium values were at the same order of magnitude as the Langmuir model. 

\begin{figure}[ht!]
    \centering
    \includegraphics[scale = 0.9]{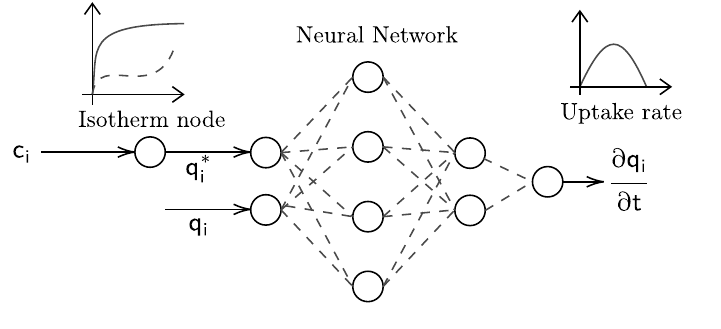}
    \caption{Schematic representation of the proposed hybrid model.}
    \label{fig:2}
\end{figure}

Here, it is important to remark that $q^*$ is assumed to be known (both structure and parameters), which contrasts with any other approach previously reported. It is a reasonable assumption as equilibrium models are usually fit separately from kinetics. Adding the task of fitting an isotherm with sorption kinetic simultaneously complicates the problem unnecessarily. \autoref{fig:2} illustrates the proposed methodology. The isotherm calculation one of the nodes in the computational graph -- the one located leftmost layer. This variable is then concatenated with the adsorbed amount and used as an input of the ANN that predicts the instantaneous uptake rate. 

The structure of the ANN is usually chosen before training it. In traditional deep learning, hyperparameter optimization is usually used to select the best architecture for a problem. However, this is rarely explored in hybrid modeling as the training procedure is significantly more computationally demanding than in traditional deep learning. Here we followed the recent literature \cite{Rackauckas2020}, where one layer and no more than 50 neurons are usually required to find the missing term in the differential equation. In the present work, one-layer ANNs with hyperbolic tangent activation were used for all cases and a varying number of neurons between 15 and 25 with grid search. Learning rates were set to 0.05 with Adaptive moment estimation (ADAM) optimizer \cite{Kingma2015} and exponential learning rate decay every 20 iterations and 0.985 drop factor over 180 iterations. After the second fit with ADAM, the quasi-newton Broyden–Fletcher–Goldfarb–Shanno (BFGS) method \cite{Nocedal2006} was employed until convergence.

\subsection{Model's numerical solution and gradient calculation}
It is important detailing the numerical solution of the Partial Differential Equation (PDE) hybrid model as it is known that the stability and speed of gradient calculation are influenced by the discretization and numerical integration method \cite{Rackauckas2020}. In this sense, detailing all the steps to solve the non-linear advection-diffusion-sorption is of utmost importance. The Partial Differential Equation (PDE) presented in \autoref{sub:hybrid} has to be solved numerically, and the resulting numerical solution is used for calculating sensitivities (gradients, Jacobians, Hessians). One approach for solving the presented model is discretizing the spatial domain and integrating the resulting Ordinary Differential Equation (ODE) in time with an appropriate numerical method -- the popular method of lines. Many choices can be made in the discretization of the spatial domain. In the literature of fixed-bed chromatography, Orthogonal Collocation on Finite Elements (OCFEM) is preferred and recommended \cite{Finlayson1974, MA1991415} as it achieves high accuracy in the presence of steep gradients in concentration profiles typically found in such problems.

This way, OCFEM is used with cubic Hermite polynomials \cite{Ganaie2014} and zeros of shifted orthogonal Legendre polynomials as collocation points. Cubic Hermite polynomials are C$^1$ continuous at the connecting nodes in OCFEM formulation, considerably reducing the required number of equations in the discretization. Thus, the solution to the PDE can be expanded as a sum of cubic Hermite basis functions as:

\begin{align}
y(u,t) &= \sum_{i = 1}^{4} a_{i + 2k - 2}(t)H_i(u), \\
H_1(u) &= (1 + 2u)(1 - u)^2, \\
H_2(u) &= u(1-u)^2h_k,\\
H_3(u) &= u^2(3-2u), \\
H_4(u) &= u^2(u-1)h_k,
\end{align}

in which,  u is the local coordinate inside each finite element, $k$ is the finite element index,$h_k$ is $k$\textsuperscript{th} finite element size, and $a(t)$ is the time-dependent coefficients solved in the ODE integration. A total of 42 evenly-spaced finite elements were used. The resulting system of ODE was then solved using a fixed-leading coefficient adaptive-order adaptive-time BDF method (FBDF) implemented in DifferentialEquations.jl with tolerances of $1\times10^{-6}$ \cite{rackauckas2017differentialequations}. FBDF presented no instability and, by far, the fastest run time for solving the discretized PDE.

Once the solution of the described model can be obtained numerically, the gradient calculation problem is formulated. As in any parameter estimation problem, a cost function has to be defined so the gradients can be calculated concerning the parameters and used in an optimizer. In this way, an L2-norm-based discrete cost function $G(\mathbf{u}, \theta)$, where $\mathbf{u}$ is the PDE solution vector, and $\theta$ is the ANNs' parameters, can be defined as:

\begin{align}
G(\mathbf{u}, \theta)  = \sum_{i = 1}^{N} || \mathbf{u}(t_i, \theta) - \mathbf{u}(t_i)^{obs}||^2,  
\end{align}

in which, $\mathbf{u_i}^{obs}$ are the vector of observations at each $i$\textsuperscript{th} point in time. In this work, $\mathbf{u_i}^{obs}$ corresponds to the liquid phase concentration at the column outlet, i.e., $c\,(x^* = 1, t_i)$. Thus, the gradient of the cost function consists in calculating $\frac{dG(u, \theta)}{d\theta}$. It means that the solution of the ODE is part of the cost function. Calculating gradients of cost functions that contain the solution of ODEs has been investigated over many years \cite{Ma2021}. Continuous sensitivity analysis with forward-mode AD is efficient when the number of equations and parameters is small. However, the sum of parameters and equations may reach hundreds in hybrid models involving discretized partial differential equations and ANNs. In this scenario, continuous or discrete adjoint sensitivity analysis is preferred. In continuous adjoint sensitivity analysis, gradient calculation for ODEs consists in solving the problem \cite{Rackauckas2020}: 

\begin{align}
\frac{d G(\mathbf{u}, \theta)}{d \theta} =  \frac{d}{d\theta} \int_{t_0}^{T}\left(\sum_{i = 1}^{N} || \mathbf{u}(t_i,\theta) - \mathbf{u_i}(t_i)^{obs}||^2\delta(t-t_i) - \mathbf{\lambda}^*\left(M\mathbf{u'} - f(\mathbf{u},\theta,t)\right)\right)dt \label{ad_1} , \\
M^*\lambda' = -\frac{df}{du}^*\lambda - \frac{dg}{du}^* \label{ad_2} \\
\frac{dg}{du}(t_i) = 2(\mathbf{u}(t_i,\theta) - \mathbf{u}(t_i)^{obs}) ,\label{ad_3}
\end{align}

in which, $M$ is the mass matrix, $\lambda$ is the Lagrange multiplier vector, and $\frac{df}{du}^*\lambda$ is the vector-jacobian product, $\delta$ is the Dirac delta function. The $*$ notation means transpose and $'$ means the time derivative.
    
The Julia SciMLSensitivity.jl library \cite{Rackauckas2020} counts with many implementations of adjoints and intermediate calculations. There are 48 ways of solving the adjoint problem, each with different performance characteristics and limitations. However, Ma et al. (2021) \cite{Ma2021} show that the quadrature adjoint with AD-jacobian vector product seeding outperforms any other method when the number of parameters + equations surpass 100. In this way, the quadrature adjoint with JIT-compiled reverse-mode AD vector-jacobian product with FBDF solver is used for solving Equations \ref{ad_1}, \ref{ad_2}, and \ref{ad_3}  for the hybrid model presented in \autoref{sub:hybrid}. It was the most performant and stable method for solving the adjoint problem resulting from a quadratic loss function and hybrid PDE.

\begin{figure}[ht!]
    \centering
    \includegraphics[scale = 0.8]{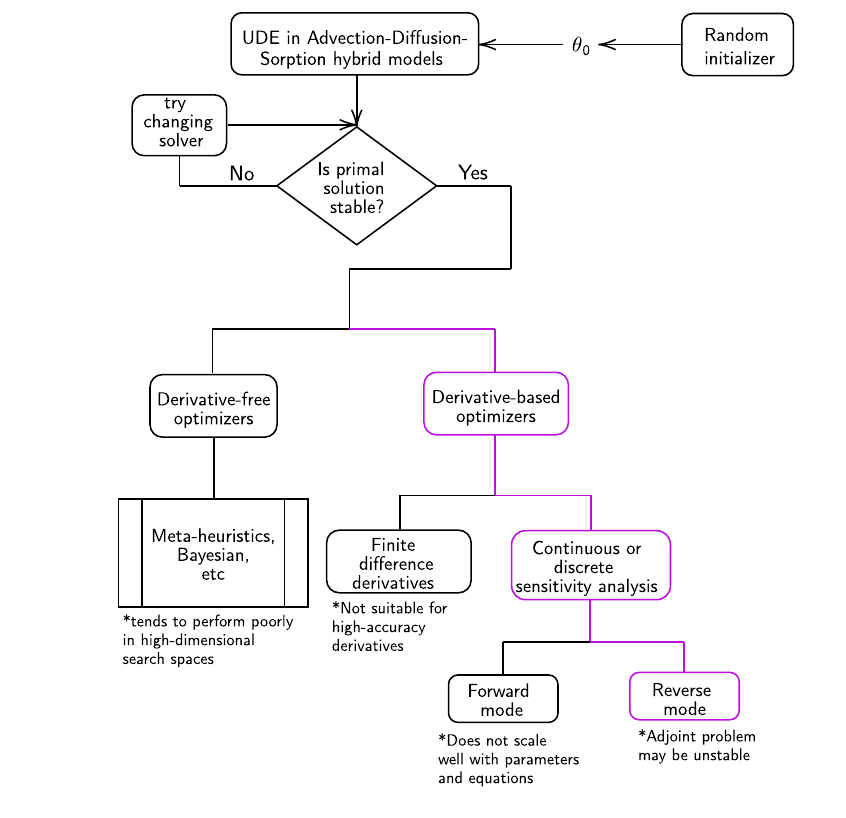}
    \caption{The decision-making process for identifying hybrid models.}
    \label{fig:diagram}
\end{figure}

\autoref{fig:diagram} illustrates the decision-making process for training hybrid models in non-linear advection-diffusion-sorption PDE systems. The lines in pink indicate the one used in this work. First, a parameter initialization algorithm has to be chosen. Then, the stability of the primal solution of the discretized PDE has to be assessed.  A decision on derivative-free or derivative-based optimizers has to be taken - if the cost function is differentiable, derivative-based optimizers are recommended. In derivative-based optimizers, adjoint sensitivity analysis is the most accurate and efficient way of calculating derivatives for large parameter space. However, the stability of the adjoint problem solution depends on the system under study. Here, a way to use the most efficient approach for fitting ANNs' weights and biases was found.

\subsection{Sparse Regression}

Sparse regression is a technique used in scientific machine learning for identifying a small subset of important variables from a more extensive set of potential predictors. Sparse regression methods aim to find a sparse solution where only a subset of the predictors is included in the final model. It results in a more straightforward and interpretable model while maintaining good predictive performance. Several methods are available for performing sparse regression, including Lasso (Least absolute shrinkage and selection operator), Ridge Regression, and SINDy(Sparse Identification of Nonlinear Dynamic systems) \cite{Brunton2016}. 

In universal differential equations and hybrid modeling, sparse regression can recover the missing interaction term in the differential equation \cite{Dandekar2020}. When trained, the ANN should resemble the missing function in the differential equation. By applying sparse regression on the trained ANN, one can recover the missing function at a computational cost significantly slower than running it directly on the original PDE. Moreover, the popular SINDy method, for instance, involves using a spline interpolation over data points. However, this approach requires a dense set of data points for accurate derivative estimates. In contrast, running sparse regression on the ANN does not require this, as it uses a trained neural network that can be continuously sampled. Assuming a trained ANN, the optimal parameter value $\hat{\theta}$ can be written as: 

\begin{align}
\hat{\theta} \leftarrow \underset{\theta}{\arg\min}\,G(\mathbf{u}, \theta) = \sum_{i = 1}^{N} || \mathbf{c}(t_i, \theta, x^* = 1) - \mathbf{c_i}^{obs}||^2/N  
\end{align}

Then, to recover the missing interaction with sparse regression, the following problem is solved:

\begin{align}
\hat{\Xi} \leftarrow \underset{\Xi}{\arg\min}\, || \textrm{ANN}(q^*(x = L, t_i), q(x = L, t_i), \hat{\theta}) - \Theta\Xi  ||_2 + \Phi||\Xi||_1 
\end{align}

Where $\Theta$ is the set of candidate basis functions (functions of $q^*$ and $q$), $\Xi$ is the parameters of the basis functions, $\Phi$ is the regularization parameter, and $t_i$ is the time when the solution of the PDE is evaluated and used in the ANN.

\section{Results}
\label{sec:results}
\subsection{UDEs' training and test sets performance}
\label{results_calibration}
This subsection shows the training and test performance of the UDE-based hybrid model proposed for all conditions displayed in \autoref{table:1}. The history of predictions and \textit{in-silico} data points are displayed as the prediction error. Moreover, the training uptake rate is also displayed and compared to the ground truth, i.e., the noise-free training set uptake rate of the mechanistic model used. 

The \autoref{table:3} displays the hyperparameters of the training process, i.e., the number of layers and neurons for each of the cases shown above. The learning rate during optimization with ADAM  started at 0.05 with an exponential decay every 20 iterations for all cases. After ADAM, the BFGS optimizer was used until convergence. 
\begin{table}[ht!]
\centering
\caption{ANNs' hyperparameters}
\label{table:3}
\begin{tabular}{cccc}
\hline
Isotherm & Kinetic      & Layers & Neurons per layer \\ \hline
Langmuir & LDF          & 1      & 17                \\
Langmuir & improved LDF & 1      & 17                \\
Langmuir & Vermeulen's  & 2      & 10, 8             \\
Sips     & LDF          & 1      & 20                \\
Sips     & improved LDF & 1      & 20                \\
Sips     & Vermeulen's  & 2      & 10, 8             \\ \hline
\end{tabular}
\end{table}

\begin{figure}[!ht]
\centering
\begin{subfigure}[!h]{0.65\textwidth}
   \includegraphics[width=1\linewidth]{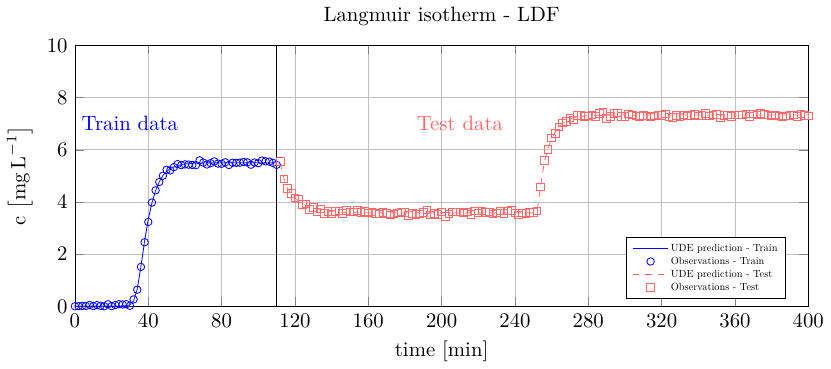}
   \caption{}
   \label{fig:2_1} 
\end{subfigure}

\begin{subfigure}[!h]{0.65\textwidth}
   \includegraphics[width=1\linewidth]{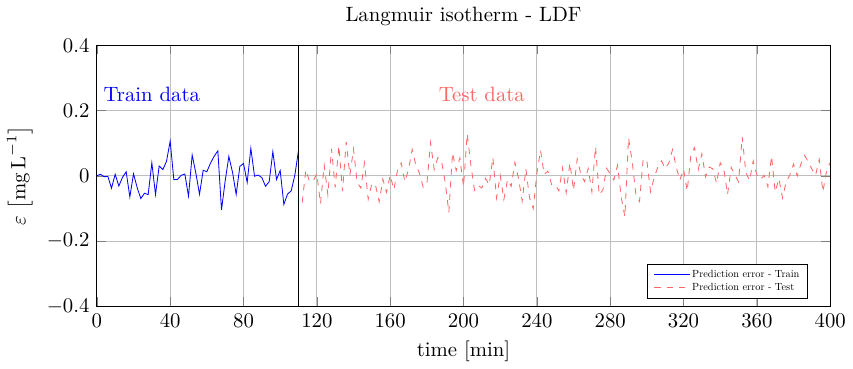}
   \caption{}
   \label{fig:2_2}
\end{subfigure}

\begin{subfigure}[!h]{0.65\textwidth}
   \includegraphics[width=1\linewidth]{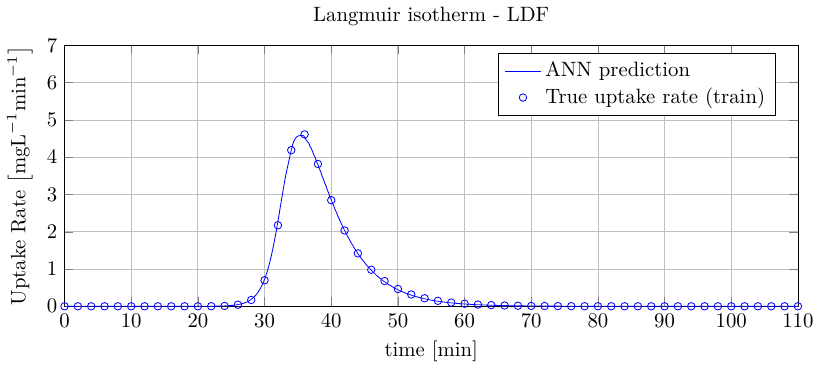}
   \caption{}
   \label{fig:2_3}
\end{subfigure}

\caption[Training and test performance]{Results for Langmuir isotherm and LDF kinetics - 0.5 min\textsuperscript{-1} sampling rate (a). Breakthrough predictions and observations for training and test sets. (b) Prediction error for training and test sets. (c) UDE prediction of uptake rate $(x^* = 1)$ compared to ground truth.}
\label{fig:4}
\end{figure}

\begin{figure}[!ht]
\centering
\begin{subfigure}[!h]{0.65\textwidth}
   \includegraphics[width=1\linewidth]{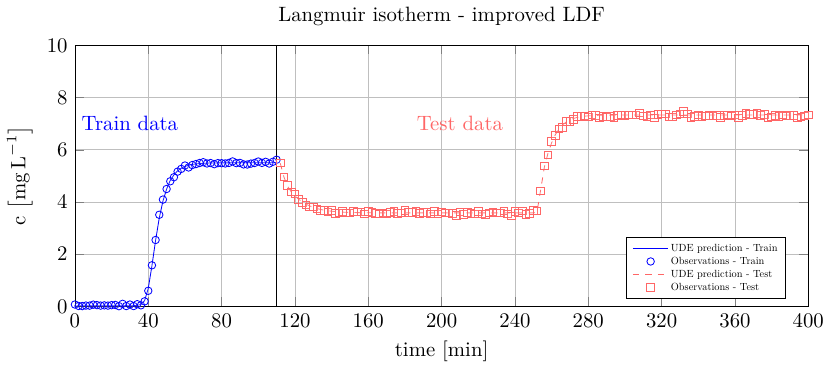}
   \caption{}
   \label{fig:3_1} 
\end{subfigure}

\begin{subfigure}[!h]{0.65\textwidth}
   \includegraphics[width=1\linewidth]{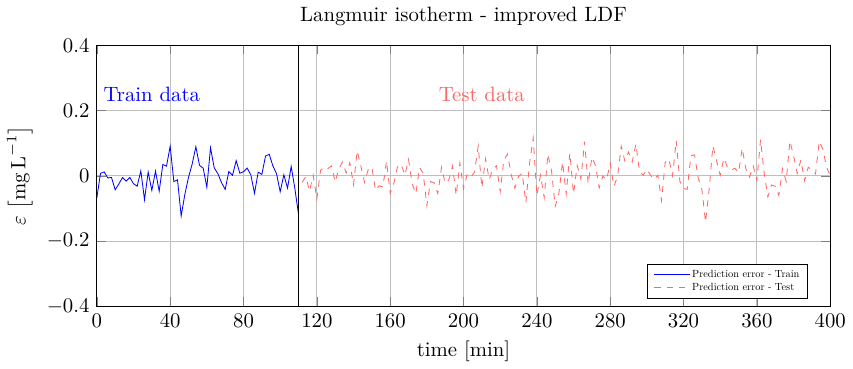}
   \caption{}
   \label{fig:3_2}
\end{subfigure}

\begin{subfigure}[!h]{0.65\textwidth}
   \includegraphics[width=1\linewidth]{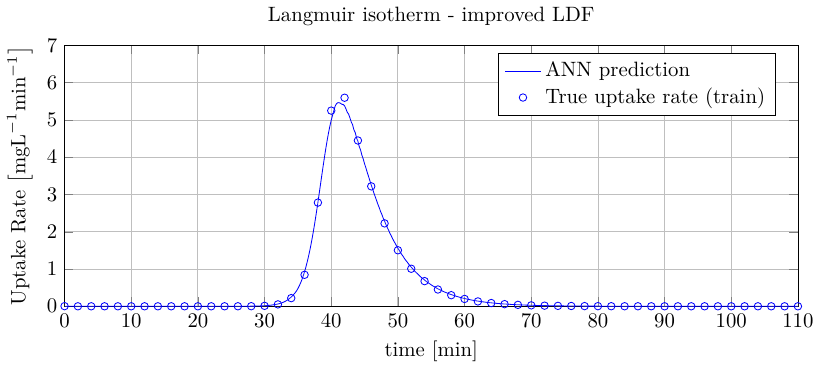}
   \caption{}
   \label{fig:3_3}
\end{subfigure}

\caption[Training and test performance]{Results for Langmuir isotherm and improved LDF kinetics - 0.5 min\textsuperscript{-1} sampling rate (a). Breakthrough predictions and observations for training and test sets. (b) Prediction error for training and test sets. (c) UDE prediction of uptake rate $(x^* = 1)$ compared to ground truth.}
\label{fig:5}
\end{figure}

\begin{figure}[!ht]
\centering
\begin{subfigure}[!h]{0.65\textwidth}
   \includegraphics[width=1\linewidth]{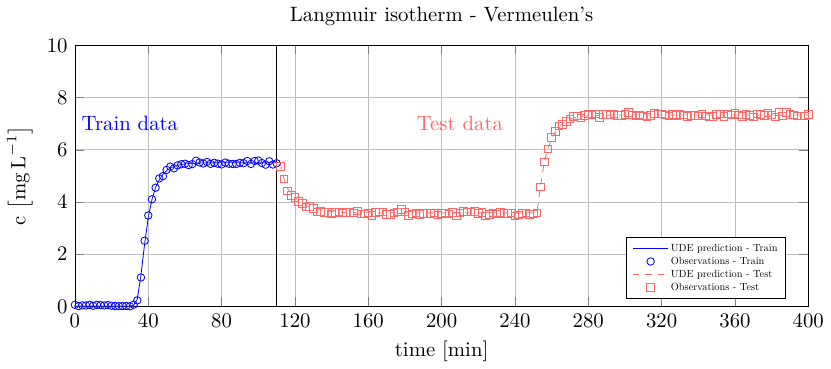}
   \caption{}
   \label{fig:4_1} 
\end{subfigure}

\begin{subfigure}[!h]{0.65\textwidth}
   \includegraphics[width=1\linewidth]{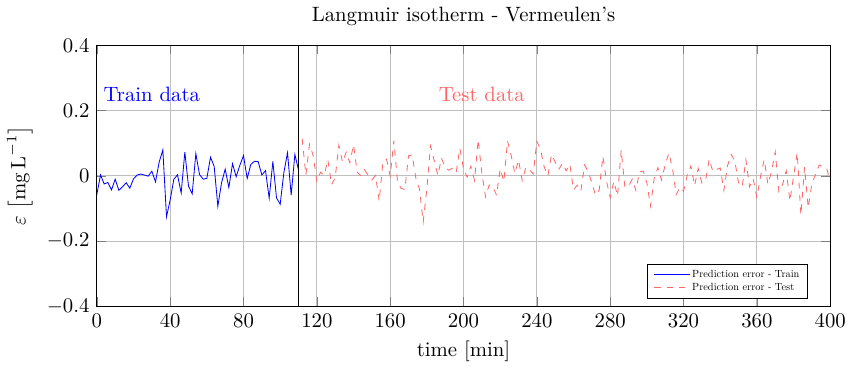}
   \caption{}
   \label{fig:4_2}
\end{subfigure}

\begin{subfigure}[!h]{0.65\textwidth}
   \includegraphics[width=1\linewidth]{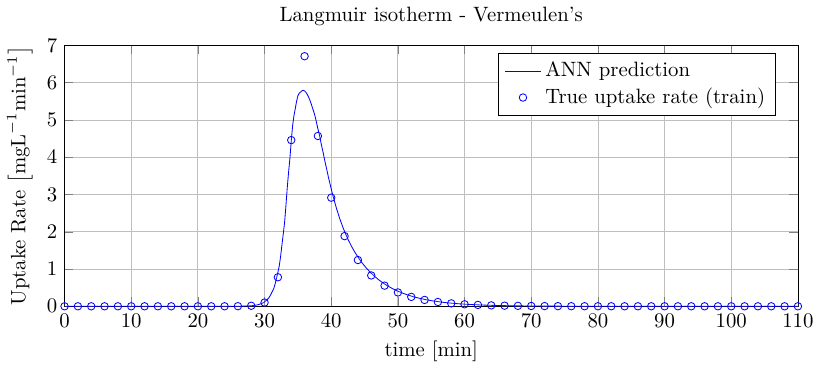}
   \caption{}
   \label{fig:4_3}
\end{subfigure}

\caption[Training and test performance]{Results for Langmuir isotherm and Vermeulen's kinetics - 0.5 min\textsuperscript{-1} sampling rate (a). Breakthrough predictions and observations for training and test sets. (b) Prediction error for training and test sets. (c) UDE prediction of uptake rate $(x^* = 1)$ compared to ground truth.}
\label{fig:6}
\end{figure}

As seen for the Langmuir isotherm and LDF kinetics in \autoref{fig:4}, the UDE approach fits breakthrough training data well, i.e., with errors compatible with simulated noise and with no apparent auto-correlation in time. Surprisingly, it also performs very well in the test set where desorption and adsorption from another steady state occur. The uptake rate is also close to the training noiseless ground-truth data. The same conclusions can be drawn for the Langmuir isotherm and improved LDF kinetics \autoref{fig:5}. However, for Vermeulen's model \autoref{fig:4_3}, the uptake rate is underestimated despite the good fitting of breakthrough data.

\begin{figure}[!ht]
\centering
\begin{subfigure}[h!]{0.65\textwidth}
   \includegraphics[width=1\linewidth]{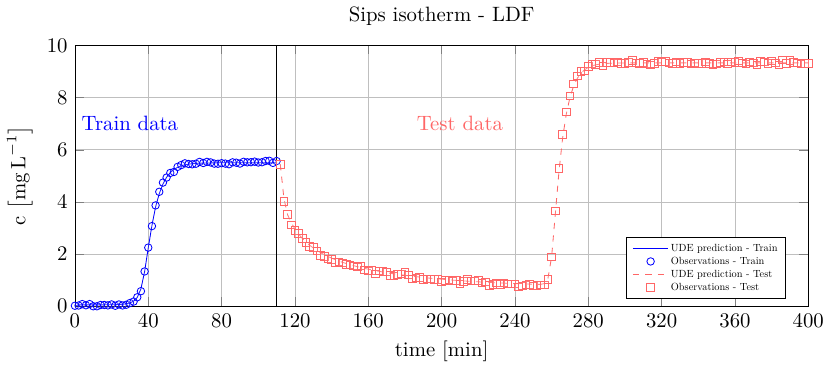}
   \caption{}
   \label{fig:5_1} 
\end{subfigure}

\begin{subfigure}[h!]{0.65\textwidth}
   \includegraphics[width=1\linewidth]{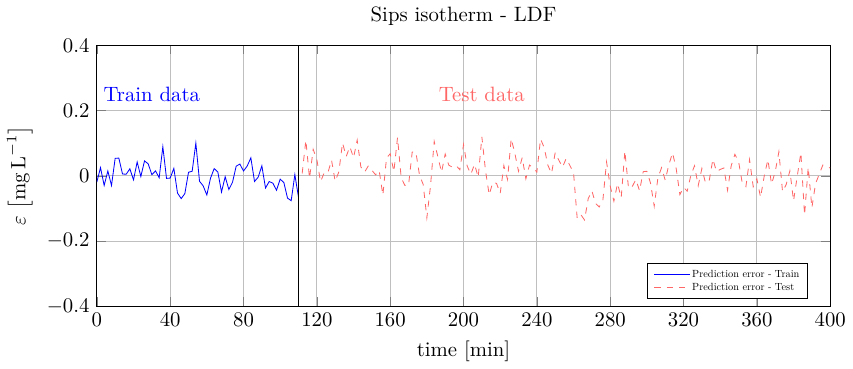}
   \caption{}
   \label{fig:5_2}
\end{subfigure}

\begin{subfigure}[h!]{0.65\textwidth}
   \includegraphics[width=1\linewidth]{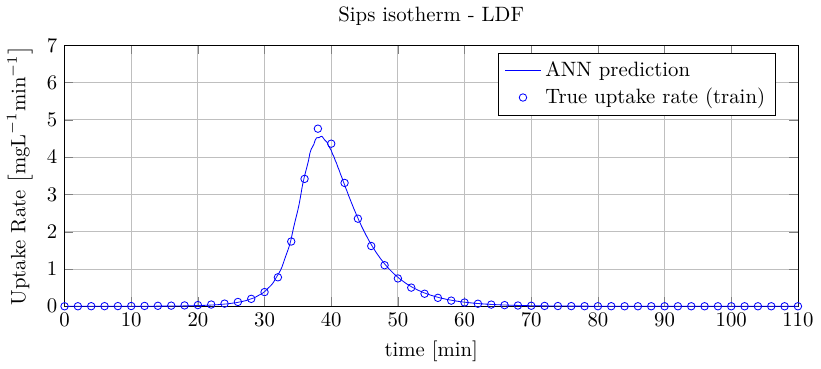}
   \caption{}
   \label{fig:5_3}
\end{subfigure}

\caption[Training and test performance]{Results for Sips isotherm and LDF kinetics - 0.5 min\textsuperscript{-1} sampling rate (a). Breakthrough predictions and observations for training and test sets. (b) Prediction error for training and test sets. (c) UDE prediction of uptake rate $(x^* = 1)$ compared to ground truth.}
\label{fig:7}
\end{figure}

\begin{figure}[!ht]
\centering
\begin{subfigure}[h!]{0.65\textwidth}
   \includegraphics[width=1\linewidth]{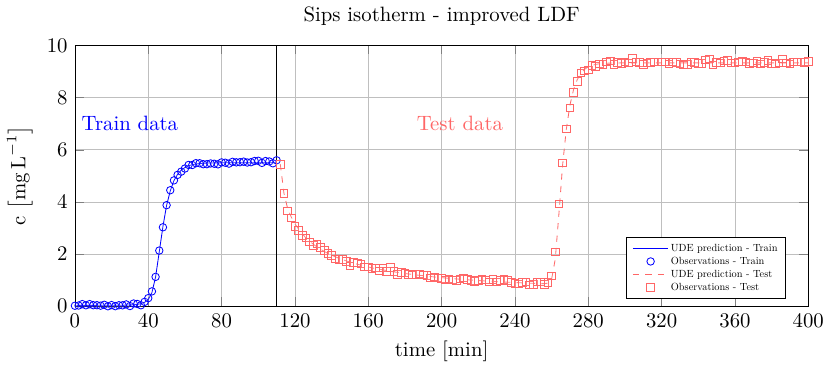}
   \caption{}
   \label{fig:6_1} 
\end{subfigure}

\begin{subfigure}[h!]{0.65\textwidth}
   \includegraphics[width=1\linewidth]{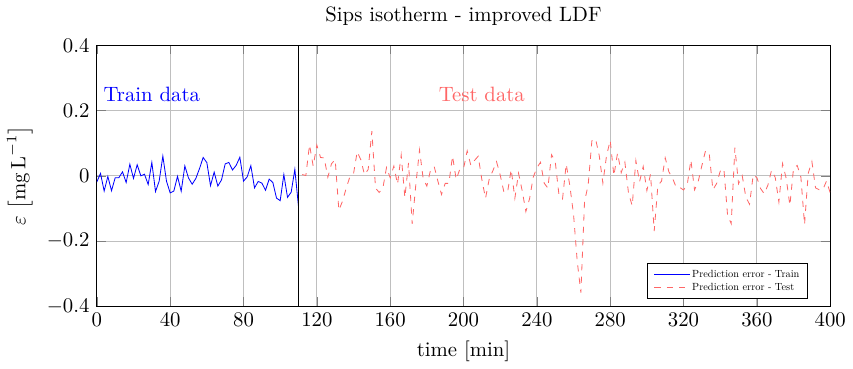}
   \caption{}
   \label{fig:6_2}
\end{subfigure}

\begin{subfigure}[h!]{0.65\textwidth}
   \includegraphics[width=1\linewidth]{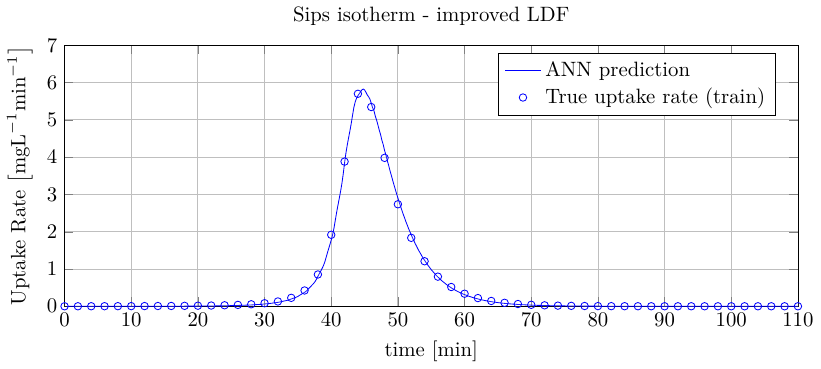}
   \caption{}
   \label{fig:6_3}
\end{subfigure}

\caption[Training and test performance]{Results for Sips isotherm and improved LDF kinetics - 0.5 min\textsuperscript{-1} sampling rate (a). Breakthrough predictions and observations for training and test sets. (b) Prediction error for training and test sets. (c) UDE prediction of uptake rate $(x^* = 1)$ compared to ground truth.}
\label{fig:8}
\end{figure}

\begin{figure}[!ht]
\centering
\begin{subfigure}[h!]{0.65\textwidth}
   \includegraphics[width=1\linewidth]{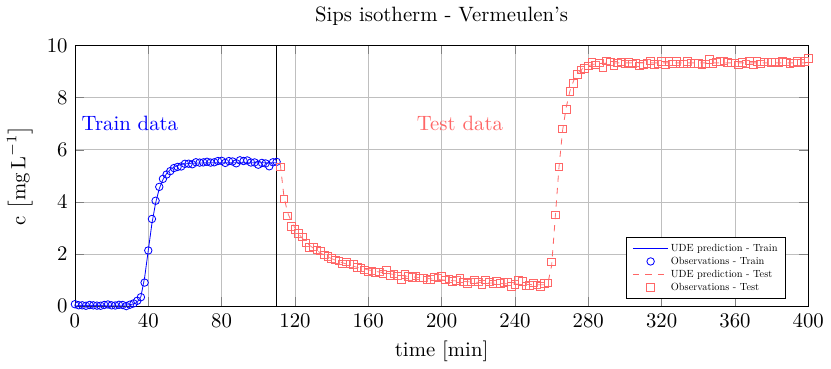}
   \caption{}
   \label{fig:7_1} 
\end{subfigure}

\begin{subfigure}[h!]{0.65\textwidth}
   \includegraphics[width=1\linewidth]{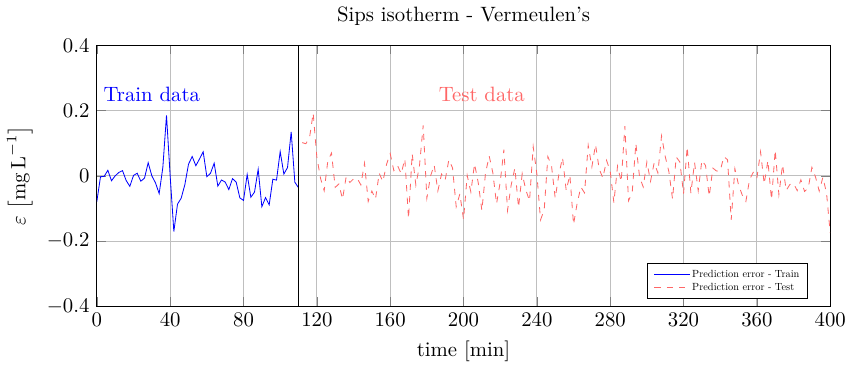}
   \caption{}
   \label{fig:7_2}
\end{subfigure}

\begin{subfigure}[h!]{0.65\textwidth}
   \includegraphics[width=1\linewidth]{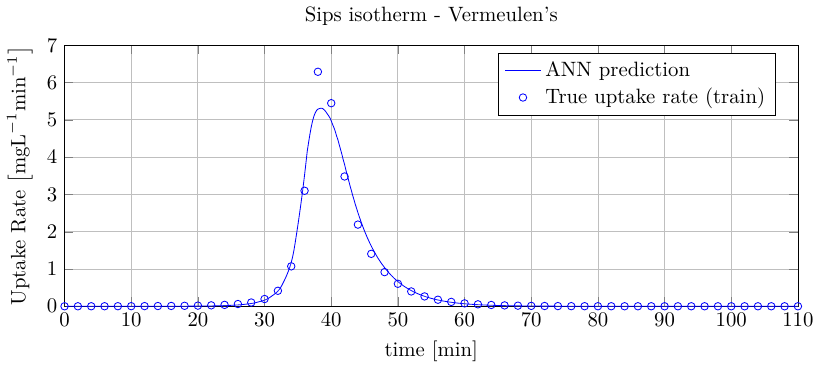}
   \caption{}
   \label{fig:7_3}
\end{subfigure}

\caption[Training and test performance]{Results for Sips isotherm and Vermeulen's kinetics - 0.5 min\textsuperscript{-1} sampling rate (a). Breakthrough predictions and observations for training and test sets. (b) Prediction error for training and test sets. (c) UDE prediction of uptake rate $(x^* = 1)$ compared to ground truth.}
\label{fig:9}
\end{figure}

As seen for the Sips isotherm and LDF kinetics in \autoref{fig:7}, the hybrid approach allows fitting breakthrough training data well, i.e., with errors compatible with simulated noise and with no apparent auto-correlation in time. Surprisingly, it also performs very well in the test set where desorption and adsorption from another steady state occur. The uptake rate is also close to the training noiseless ground-truth data. The same conclusions can be drawn for the Sips isotherm and improved LDF kinetics \autoref{fig:8}, except for the last adsorption experiment, where an error of about -0.3 is observed at the beginning of the breakthrough curve. It's an acceptable error as the system was more stressed by higher steps in the feed concentration compared to the Langmuir isotherm case. For Vermeulen's model, \autoref{fig:9}, the uptake rate is again underestimated despite the good fitting of breakthrough data.

\clearpage

\subsection{Sparse Regression}

As seen in \autoref{results_calibration}, the fitted ANN showed a reasonable train and test set performance for predicting breakthrough data. The uptake rates are also close to the ground truth. However, the neural network outputs are hard to interpret. In this sense, sparse regression can be used to understand the missing functions that the ANN approximates. Here, a set of basis polynomials up to sixth order are used to carry sparse regression using the SINDy algorithm, i.e., $\Theta \in q^{*i} \times q^j$ $i,j \in (0, 1, 2, 3, 4, 5, 6)$. That results in 49 basis to be combined in regression. The uptake rate at three different positions in the bed was used for carrying out the regression. 

As sparse regression requires tuning the regularization parameter $\Phi$, a grid search on $\Phi$s in $\textrm{log}_{10}$ space from -3.0 to 0.0 with 0.05 step was carried using ADMM algorithm in DataDrivenDiffEq.jl \cite{datadrivendiffeq} (an implementation of Lasso - Least absolute shrinkage and selection operator). The Bayesian Information Criterion (BIC) is compared for each $\Phi$ and used to decide which model is the best -- the lower the BIC, the higher the chance that a model explains the observations. BIC is a metric that combines the model's predictive capacity (squared residuals) and complexity (number of parameters). Here it is tested if this approach allows reconstructing the known kinetics originating the data, i.e., LDF, improved LDF, and Vermeulen's model.  After running the sparse regression on the ANN, the parameters are readjusted using BFGS optimizer over 1000 iterations.

It is important to note that the need to carry sparse regression on the ANN outputs appears when test set (extrapolation) performance is poor, or ANNs' interpretability is desired. However, it was demonstrated that the ANNs performed surprisingly well in extrapolation tasks. In this way, the attempt to recover interaction terms in the current work serves only the purpose of giving interpretability to ANNs' outputs.  

\subsubsection{Langmuir Isotherm}
\begin{figure}[htpb!]
\centering
\includegraphics[width=0.95\linewidth]{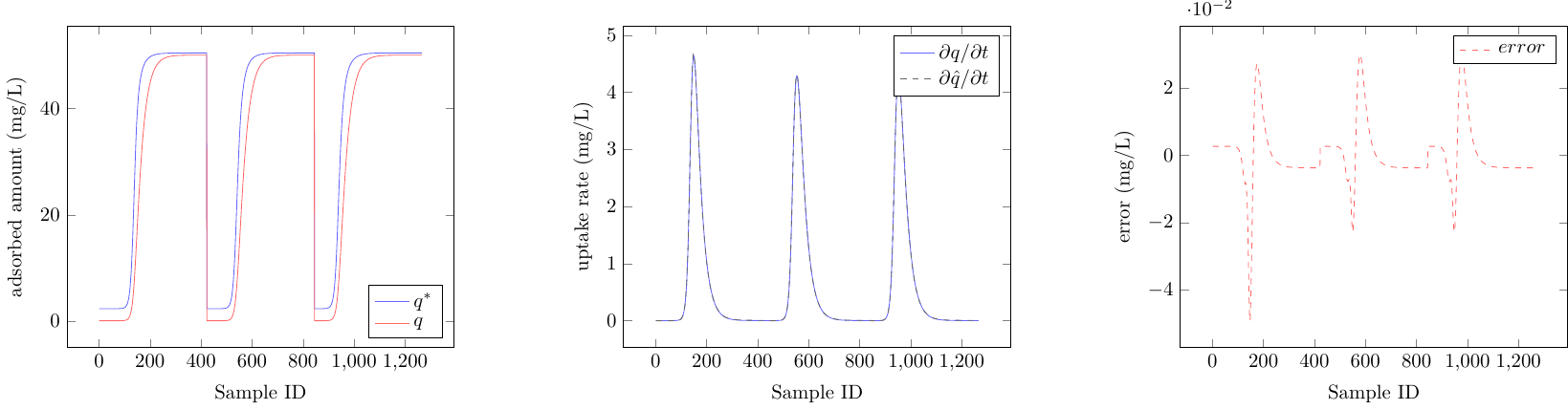}
\caption{Sparse regression results for the best number of active terms (3) - Langmuir isotherm + LDF kinetics at three positions in the bed.}
\label{fig:10}
\end{figure}

\begin{figure}[htpb!]
\centering
\includegraphics[width=0.65\linewidth]{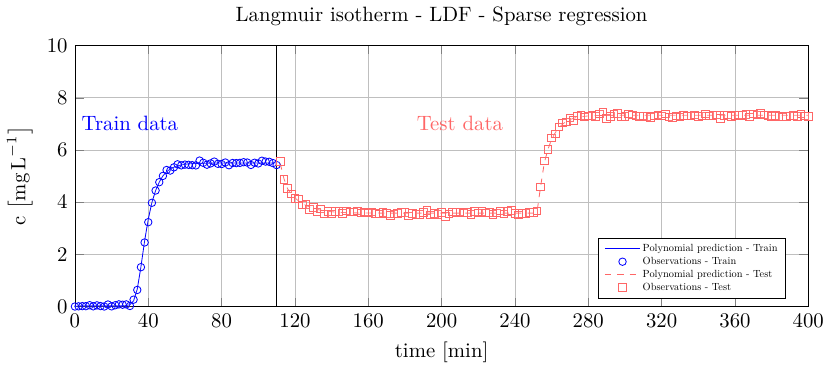}
\caption{Results for Langmuir isotherm and LDF kinetics -  Breakthrough
predictions and observations for training and test sets using the identified polynomial.}
\label{fig:11}
\end{figure}

For the Langmuir isotherm and LDF kinetics, 3 was the number of active terms that minimized the Bayesian Information Criterion (BIC).  The uptake rate at three positions in the bed was used for the regression to improve accuracy. The obtained equation was $p_1 + p_2q + p_3(q^*)$ where $p_1 = -0.535, p_2 = -0.225 , p_3 = 0.234$. \autoref{fig:10} shows that a good fitting is achieved and that the dominant coefficients are the order 1 exponent, which agrees with the original interaction term \autoref{ap_1}.  After running sparse regression on the ANN, the obtained polynomial can be used in the original PDE for carrying predictions. \autoref{fig:11} shows that the breakthroughs are well predicted with this 3-term polynomial in the kinetics.   

\begin{figure}[htpb!]
\centering
\includegraphics[width=0.95\linewidth]{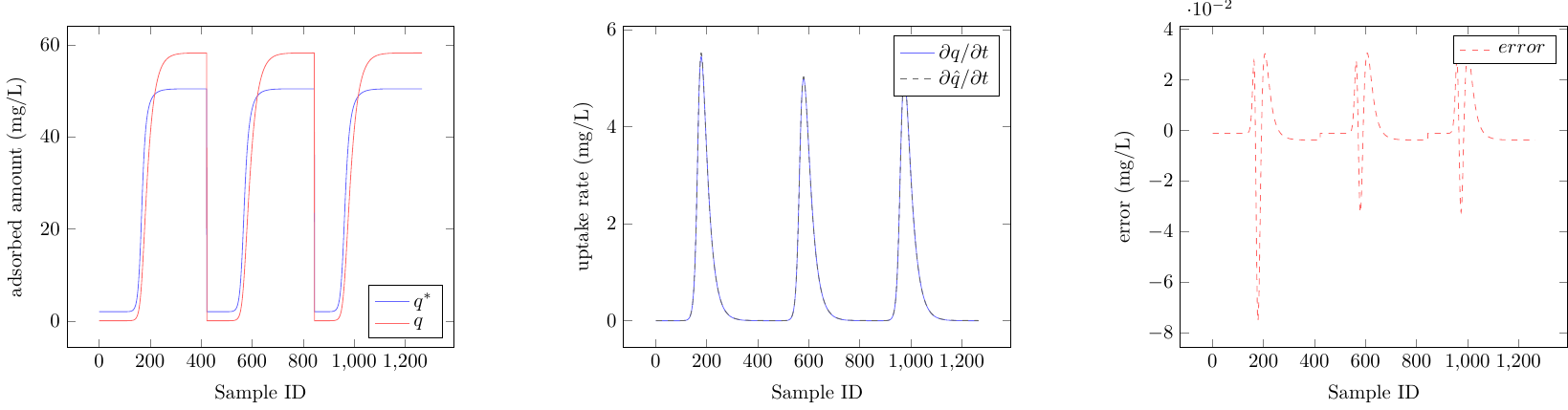}
\caption{Sparse regression results for the best number of active terms (4) - Langmuir isotherm + improved LDF kinetics at three positions in the bed.}
\label{fig:12}
\end{figure}

\begin{figure}[ht!]
\centering
\includegraphics[width=0.65\linewidth]{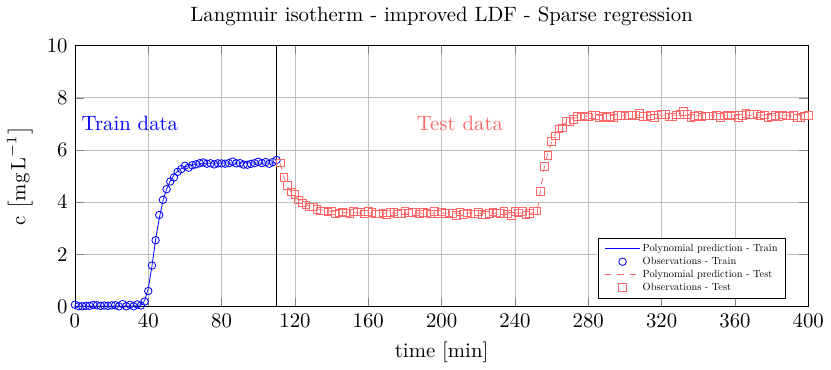}
\caption{Results for Langmuir isotherm and improved LDF kinetics -  Breakthrough
predictions and observations for training and test sets using the identified polynomial.}
\label{fig:13}
\end{figure}

For the Langmuir isotherm and improved LDF kinetics, 4 was the number of active terms that minimized the Bayesian Information Criterion (BIC).  The uptake rate at three positions in the bed was used for the regression to improve accuracy. The obtained equation was $p_1 + p_2q + p_3(q^*) + p_4qq^*$ where $p_1 = -0.549, p_2 = -0.221, p_3 = 0.278, p_4 = -0.000212$. \autoref{fig:12} shows that a good fitting is achieved. \autoref{fig:13} shows that the breakthroughs are well predicted with this 4-terms polynomial in the kinetics.   

\begin{figure}[ht!]
\centering
\includegraphics[width=0.95\linewidth]{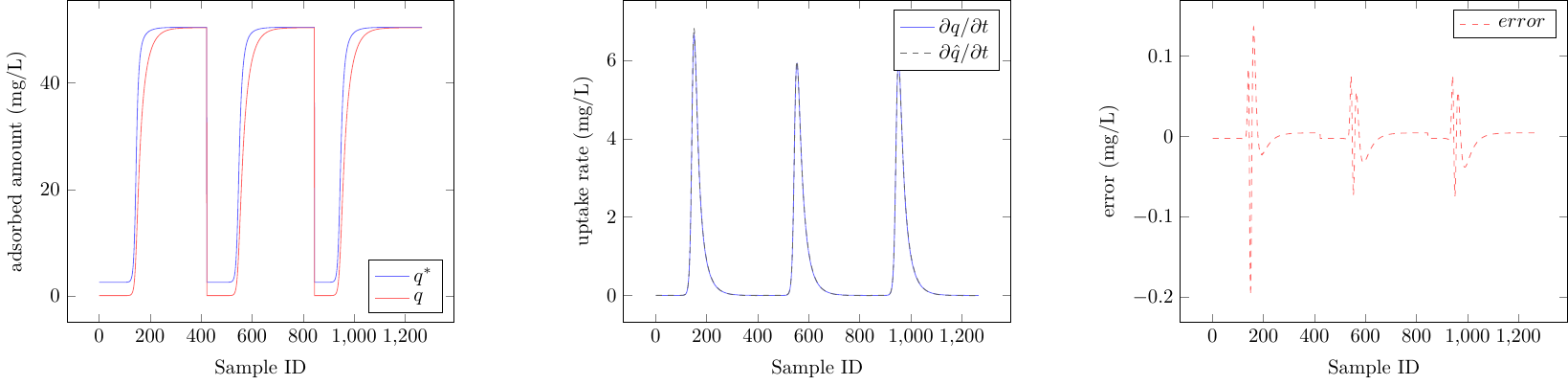}
\caption{Sparse regression results for the best number of active terms (6) - Langmuir isotherm + Vermeulen's kinetics at three positions in the bed. }
\label{fig:131}
\end{figure}

\begin{figure}[ht!]
\centering
\includegraphics[width=0.65\linewidth]{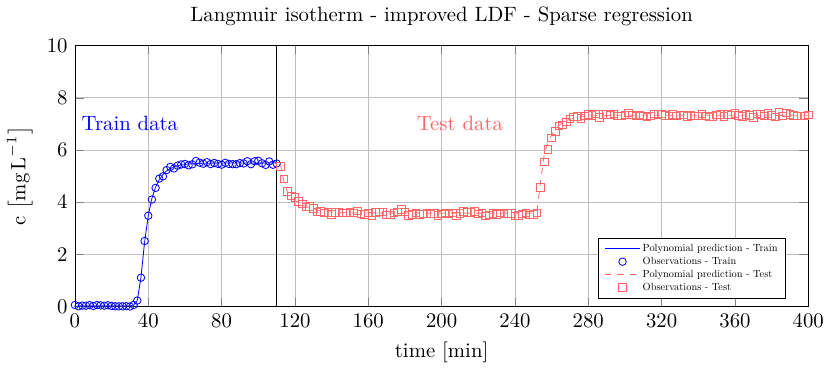}
\caption{Results for Langmuir isotherm and Vermeulen's kinetics -  Breakthrough
predictions and observations for training and test sets using the identified polynomial.}
\label{fig:132}
\end{figure}

For the Langmuir isotherm and Vermeulen's kinetics, 6 was the number of active terms that minimized the Bayesian Information Criterion (BIC). The obtained equation was $-1.253 -0.429q + 0.00338(q^2) + 0.5064q^* - 0.00236(q^{*2}) -0.00207q^*$. \autoref{fig:131} shows that a good fitting is achieved. \autoref{fig:132} shows that training and test breakthroughs are well predicted with this 6-term polynomial in the kinetics. However, the polynomial interpretation is not straightforward due to the $q$ term in the denominator in \autoref{ap_2}.


\subsubsection{Sips Isotherm}
\begin{figure}[!ht]
\centering
\includegraphics[width=0.95\linewidth]{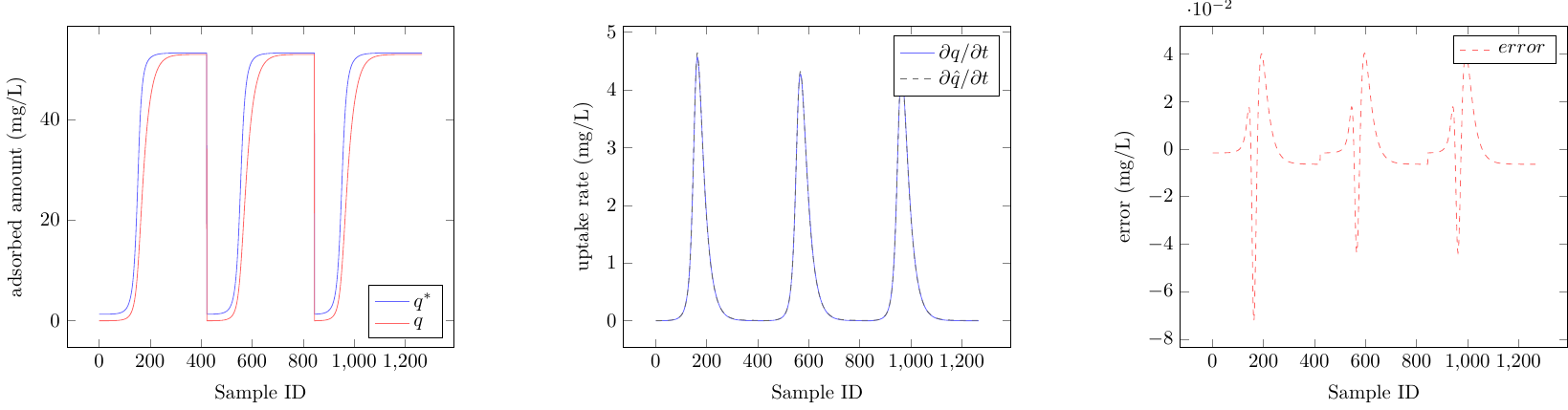}
\caption{Sparse regression results for the best number of active terms (3) - Sips isotherm + LDF kinetics at three positions in the bed.}
\label{fig:14}
\end{figure}

\begin{figure}[!ht]
\centering
\includegraphics[width=0.65\linewidth]{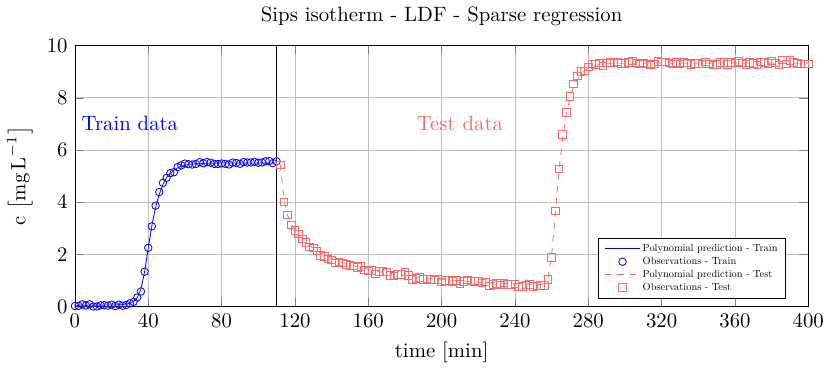}
\caption{Results for Sips isotherm and LDF kinetics -  Breakthrough
predictions and observations for training and test sets using the identified polynomial.}
\label{fig:15}
\end{figure}

For the Sips isotherm and LDF kinetics, 3 was the number of active terms that minimized the Bayesian Information Criterion (BIC).  The uptake rate at three positions in the bed was used for the regression to improve accuracy. The obtained equation was $p_1 + p_2q + p_3(q^*)$ where $p_1 = -0.275, p_2 = -0.210 , p_3 = 0.214$. \autoref{fig:14} shows that a good fitting is achieved and that the dominant coefficients are the order 1 exponent, which agrees with the original interaction term \autoref{ap_1}.  After running sparse regression on the ANN, the obtained polynomial can be used in the original PDE for carrying predictions. \autoref{fig:15} shows that the breakthroughs are well predicted with this 3-terms polynomial in the kinetics.

\begin{figure}[!ht]
\centering
\includegraphics[width=0.95\linewidth]{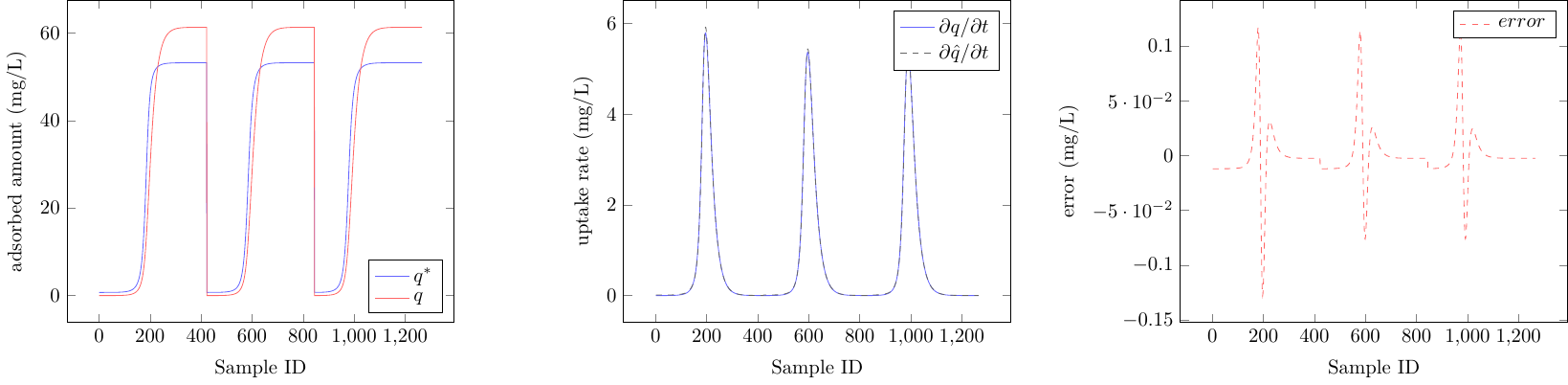}
\caption{Sparse regression results for the best number of active terms (4) - Sips isotherm + improved LDF kinetics at three positions in the bed.}
\label{fig:16}
\end{figure}

\begin{figure}[!ht]
\centering
\includegraphics[width=0.65\linewidth]{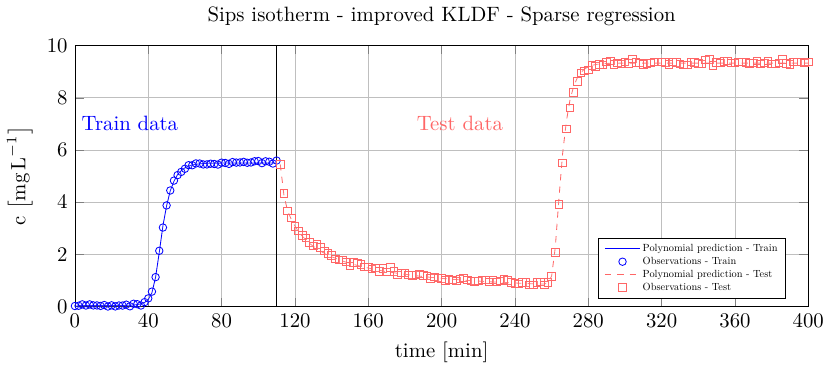}
\caption{Results for Sips isotherm and improved LDF kinetics -  Breakthrough
predictions and observations for training and test sets using the identified polynomial.}
\label{fig:17}
\end{figure}

For the Sips isotherm and improved LDF kinetics, 4 was the number of active terms that minimized the Bayesian Information Criterion (BIC). The obtained equation was $p_1 + p_2q + p_3(q^2) + p_4(q^*)$ where $p_1 = -0.183, p_2 = -0.215 , p_3 = -0.000294, p_4 = 0.272$. \autoref{fig:16} shows that a good fitting is achieved. In this case, interpreting the obtained polynomial is not straightforward due to an exponential term in the original interaction (\autoref{ap_3}). \autoref{fig:17} shows that the breakthroughs are well predicted with this 4-term polynomial in the kinetics.

\begin{figure}[!ht]
\centering
\includegraphics[width=0.95\linewidth]{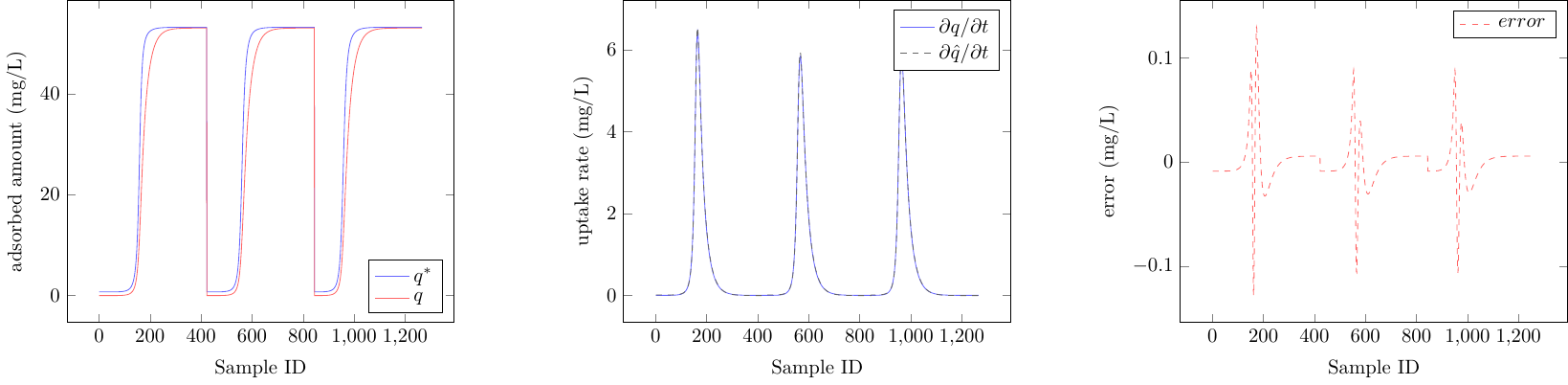}
\caption{Sparse regression results for the best number of active terms (6) - Sips isotherm + Vermeulen's kinetics at three positions in the bed. }
\label{fig:18}
\end{figure}

\begin{figure}[!ht]
\centering
\includegraphics[width=0.65\linewidth]{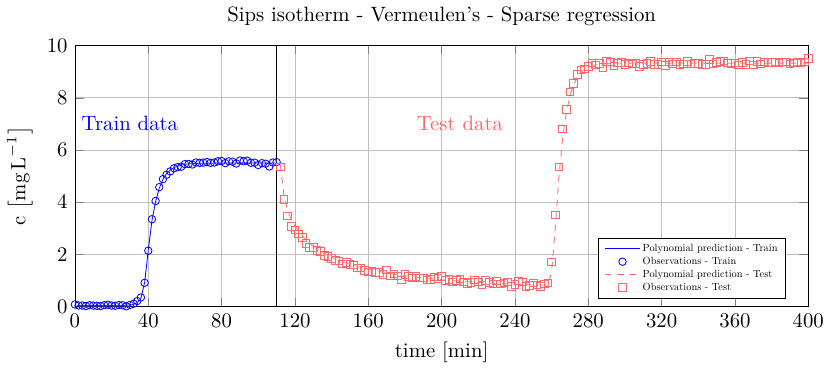}
\caption{Results for Sips isotherm and Vermeulen's kinetics -  Breakthrough
predictions and observations for training and test sets using the identified polynomial.}
\label{fig:19}
\end{figure}

For the Sips isotherm and Vermeulen's kinetics, 6 was the number of active terms that minimized the Bayesian Information Criterion (BIC). The obtained equation was $p_1 + p_2q + p_3(q^2) + p_4q^* + p_6(q^{*3}) + p_5q(q^{*2})$ where $p_1 = -0.309, p_2 = -0.449 , p_3 = 0.00037, p_4 = 0.438, p_5 = 6.214 \times 10^{-5}, p_6 = -6.326 \times 10^{-5}$. \autoref{fig:18} shows that a good fitting is achieved. \autoref{fig:19} shows that training and test breakthroughs are well predicted with this 6-term polynomial in the kinetics. Once again, the polynomial interpretation is not straightforward due to the $q$ term in the denominator in \autoref{ap_2}.


\autoref{tab:sparse} summarizes the results of sparse regression for each case. It can be noted that the obtained polynomial resembles the original kinetics for the LDF kinetics (Langmuir and Sips isotherms). The train and test set predictions are very close to the noisy observations.  

\begin{table}[!ht]
\caption{Sparse regression learned polynomials.}
\begin{tabular}{cccc}
\hline
Isotherm & Kinetic      & True kinetics                                     & Learned kinetics                                                                                                                                         \\ \hline
Langmuir & LDF          & $0.22q^* - 0.22q$                                & $-0.535 -0.225q + 0.234(q^*)$                                                                                                                         \\
Langmuir & improved LDF & $0.22(q^* + 0.2789q^*e^{{\frac{-q}{2q^*}}} - q)$ & $-0.549 -0.221q + 0.278(q^*) -0.000212qq^*$                                                                                                         \\
Langmuir & Vermeulen's  & $0.22\frac{q^{*2} - q^2}{2.0q}$                    & \begin{tabular}[c]{@{}c@{}} $-1.253 -0.429q + 0.0034(q^2) + 0.506q^*$ \\ $-0.0023(q^{*2}) -0.0021qq^*$  \end{tabular}                                                                                                                                                 \\
Sips     & LDF          & $0.22q^* - 0.22q$                                & $-0.275  -0.210q + 0.214(q^*)$                                                                                                                         \\
Sips     & improved LDF & $0.22(q^* + 0.2789q^*e^{{\frac{-q}{2q^*}}} - q)$ & $ -0.183 -0.215q  -0.00029(q^2) + 0.272(q^*)$                                                                                                        \\
Sips     & Vermeulen's  & $0.22\frac{q^{*2} - q^2}{2.0q}$                    & \begin{tabular}[c]{@{}c@{}}$-0.309 -0.449q + 0.00037(q^2) + 0.438q^*$ \\ $ -6.326 \times 10^{-5}(q^{*3}) + 6.214 \times 10^{-5}q(q^{*2})$\end{tabular} \\ \hline
\end{tabular}
\label{tab:sparse}
\end{table}

\clearpage
\subsection{Symbolic Regression}
The trained ANNs' outputs were also regressed by using symbolic regression. Symbolic regression is a technique that aims to uncover a model's structure and parameters in a single step. This process involves searching through a space of mathematical expressions to find the best representation of the data rather than relying on a pre-defined model structure. One of the most common approaches to solving the symbolic regression problem is genetic programming (GP). This method involves creating a system of arithmetic operators (addition, subtraction, multiplication, and division) and functions (identity, logarithmic, exponential, trigonometric, and polynomial functions) to generate an expression tree. The expression tree represents a mathematical formula, and the goal is to optimize the order of the operators in the tree to achieve the best fit for the data. Several techniques of GP have been used to carry out symbolic regression. In this particular case, the approach used was the scoring method proposed by Cranmer et al. (2020) \cite{cranmer2020discovering}, implemented in Julia language using the SymbolicRegression.jl library \cite{pysr}. The algorithm was run for over 30 iterations with a population size of 30. A Pareto front of polynomials was generated, and the polynomial with the lowest mean squared error was chosen as the optimal solution.

One of the advantages of symbolic regression compared to sparse regression is that it does not require tuning the sparsity parameter $\Phi$ -- a parameter with considerable influence in sparse regression results. \autoref{tab:symbolic} shows the obtained polynomials using symbolic regression. Learned polynomials are similar to the ones obtained via sparse regression - differences appeared in structure, but residuals were at the same order of magnitude ($10^{-2}$). In LDF kinetics and Sips isotherm, the obtained structure is identical to the original, with very close coefficients. Expressions in symbolic regression, in general, are simpler (with fewer terms) than those obtained with sparse regression. Moreover, symbolic regression did not require grid search through sparsity parameter space and a coefficient refinement with BFGS as in sparse regression.

\begin{table}[htpb!]
\caption{Symbolic regression learned polynomials.}
\begin{tabular}{cccc}
\hline
Isotherm & Kinetic      & True kinetics                                     & Learned kinetics                                                                                                                                         \\ \hline
Langmuir & LDF          & $0.22q^* - 0.22q$                                & $-0.535 -0.225q + 0.234(q^*)$                                                                                                                         \\
Langmuir & improved LDF & $0.22(q^* + 0.2789q^*e^{{\frac{-q}{2q^*}}} - q)$ & $-0.554 -0.234q + 0.281(q^*)$                                                                                                         \\
Langmuir & Vermeulen's  & $0.22\frac{q^{*2} - q^2}{2.0q}$                    & \begin{tabular}[c]{@{}c@{}} $-0.6098 + 0.0122q + 0.263q^*$ \\ $ -0.00526qq^*$  \end{tabular}                                                                                                                                                 \\
Sips     & LDF          & $0.22q^* - 0.22q$                                & $0.198q^* -0.200q$                                                                                                                         \\
Sips     & improved LDF & $0.22(q^* + 0.2789q^*e^{{\frac{-q}{2q^*}}} - q)$ & $ 0.277q^* - 0.241q$                                                                                                        \\
Sips     & Vermeulen's  & $0.22\frac{q^{*2} - q^2}{2.0q}$                    & $- 0.003557q^{*2} - 0.216q + 0.395q^*$ \\ \hline
\end{tabular}
\label{tab:symbolic}
\end{table}

These results demonstrate that retrieving the correct interaction term from noisy observations in such problems is not trivial, despite the high-accuracy fitting (\autoref{fig:4} to \autoref{fig:9}). Most of the obtained polynomials resemble the true kinetic models that produced the data. The obtained polynomials, specifically those from the improved LDF and Vermeulen's cases, cannot be directly compared with the original terms. However, a Taylor expansion analysis of the mechanistic model results, as shown in the equations in \autoref{append}, can demonstrate why a polynomial with positive integer coefficients (i.e., without $1/q$ and $\exp(-0.5q^/q)$ basis) can approximate the true kinetics for such cases. For example, a second-order Taylor expansion was conducted for Vermeulen's model and Langmuir isotherm at sample ID = 50, around $q^* = 50.13$ and $q = 48.11$. This resulted in the following equation:

\begin{equation}
0.22\frac{q^{*2} - q^2}{2.0q}(50.13, 48.11) \approx 0.454 + 0.229q^* - 0.229q + 0.002286q^{*2} - 0.004764q^*q  + 0.002482q^2 + \mathcal{O}(\rVert x^3 \rVert)    
\end{equation}

\begin{figure}[!ht]
    \centering
    \includegraphics[scale = 0.65]{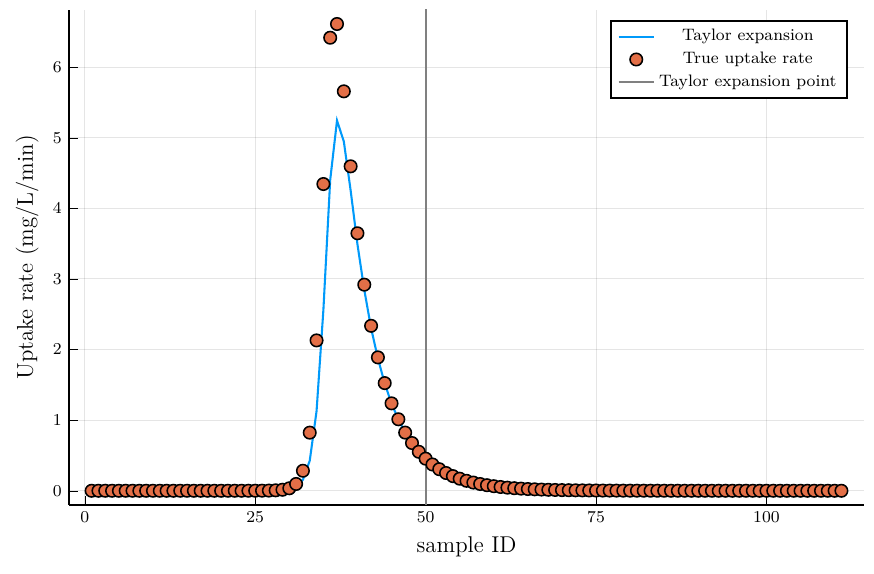}
    \caption{Taylor expansion of uptake rate for Vermeulen's model and Langmuir isotherm - Mechanistic model}
    \label{fig:taylor_quad}
\end{figure}

\autoref{fig:taylor_quad} shows how well a simple second-order Taylor expansion around sample ID = 50 can capture the entire behavior of the uptake rate for points far away from the expansion point. It can explain why polynomials with positive integer exponents found in sparse and symbolic regression explain the observations well. 

A similar analysis can be done for improved LDF kinetics. A first-order Taylor expansion at sample ID 50 ($q^* = 49.23$ and $q = 49.22$) of the mechanistic model results gives:

\begin{equation}
0.22(q^* + 0.2789q^*e^{{\frac{-q}{2q^*}}} - q)(49.23, 49.22) \approx 1.834 + 0.275q^* - 0.238q + \mathcal{O}(\rVert x^2 \rVert)    
\end{equation}

\begin{figure}[!ht]
    \centering
    \includegraphics[scale = 0.65]{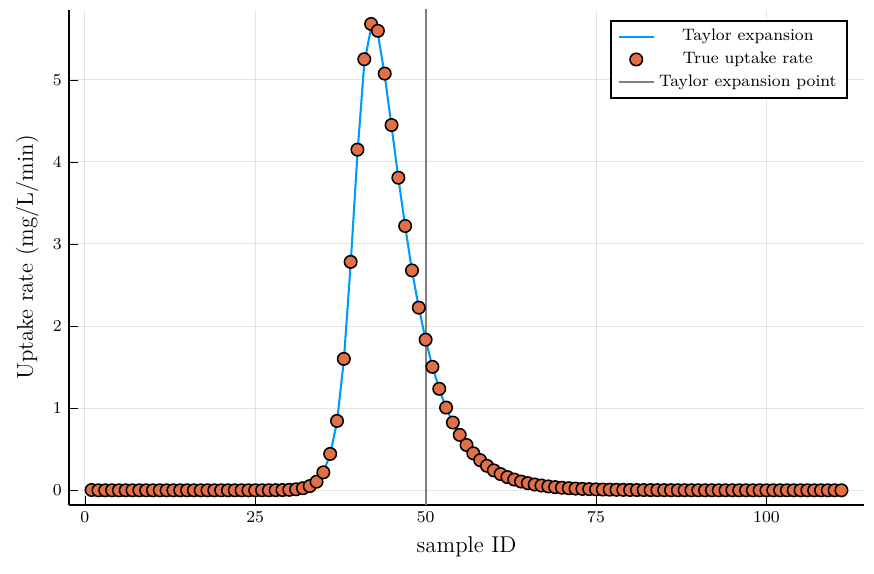}
    \caption{Taylor expansion of uptake rate for improved LDF model and Langmuir isotherm - Mechanistic model}
    \label{fig:taylor_improved}
\end{figure}

Once again, by plotting the Taylor polynomial against the true uptake rate (\autoref{fig:taylor_improved}), it is possible to see how well the points match and explain why both symbolic and sparse regressions succeed in explaining observations with such structure (polynomials with positive integer exponents).

To conclude, we remark that combining the uptake rate plots with sparse/symbolic regression produces an efficient tool for investigating sorption mass transfer kinetics in non-linear advection-diffusion-sorption systems. The difference in the functional representation learned by Lasso/symbolic regression compared to the original functions could be attributed (i) to noisy data and (ii) the fact that the two functional representations are numerically equivalent in this range of values and response shape. 

\clearpage
\section{Conclusions}

The present work investigated the numerical aspects of  hybrid non-linear advection-diffusion-sorption PDE problems. In this sense, we proposed a framework for efficiently training ANN-based hybrid non-linear advection-diffusion-sorption PDE problems using gradient-based optimizers via (i) continuous adjoint sensitivity analysis, (ii) orthogonal collocation on finite element PDE discretization scheme and an (iii) fixed-leading coefficient adaptive-order adaptive-time BDF method ODE integrator.  Another contribution of the present work was the investigation of the interpretability of ANNs' prediction through sparse and symbolic regression. To demonstrate the contributions above, an in-silico data set was used. The data set was built by simulating a hypothetical single-component non-linear advection-diffusion-sorption system with two different adsorption equilibrium (Langmuir and Sips) and three kinetic models -- Linear Driving Force (LDF), Vermeulen’s model, and improved LDF. The results show that the ANN can learn adsorption uptake from single noisy breakthroughs at column outlet and extrapolate in adsorption and desorption well.

Moreover, by carrying sparse and symbolic regression on ANNs' outputs, polynomials resembling the true kinetic models that produced the data were obtained - especially for the LDF kinetics. In the cases of non-trivial resemblance,i.e., improved LDF and Vermeulen's, a polynomial with high predictive power was obtained, and its appearance was elucidated with Taylor expansion analysis. It was demonstrated that the proposed framework could be used for kinetic model discovery from single-position noisy breakthrough data in advection-diffusion-sorption PDEs.   

\section{Code Availability}
All codes are publicly available at \url{https://github.com/viniviena/ude_chromatography/tree/master/UDE_paper_chromatography}

\section{Acknowledgements}

This research was supported by the doctoral Grant (reference PRT/BD/152850/2021) financially supported by the Portuguese Foundation for Science and Technology (FCT), LA/P/0045/2020 (ALiCE), UIDB/50020/2020 and UIDP/50020/2020 (LSRE-LCM) and with funds from FCT/MCTES (PIDDAC) and State Budgets under MIT Portugal Program.

\clearpage
\appendix
\section{A mechanistic model for the \textit{in-silico} data set}
\label{append}

A fully-mechanistic model for column chromatography model was implemented to create the data set, also using the method of lines with cubic Hermite polynomial in OCFEM. 60 evenly-spaced finite elements were used in the discretization, and the ODE integrated with the FBDF method with $1\times10^{-8}$ absolute and relative tolerances. The mechanistic model assumes a plug flow, negligible pressure drop, isothermal operation, and uniform distribution radial with no variation in the superficial flow velocity throughout the bed.

The adsorption uptake rate, however, can take three forms: LDF, quadratic, and improved LDF. The LDF assumes that the uptake rate of the adsorbate is proportional to the difference between the specie at the outer surface of the particle (equilibrium adsorption capacity) and its average concentration with the particle. The quadratic assumes that the uptake rate is proportional to the difference of concentrations squared. The improved LDF is a proposed modification where the uptake rate is corrected for low solid phase concentrations. In this way, the fully mechanistic model can be written as:
\setcounter{equation}{0}
\renewcommand{\theequation}{\thesection.\arabic{equation}}
\begin{align*}
\frac{\partial c}{\partial t^*} = -\frac{1-\varepsilon}{\varepsilon}g_i(q, q^*)\tau_s - \frac{\partial c}{\partial x^*} + \frac{1}{Pe}\frac{\partial c^2}{\partial x^{*2}} \\
\frac{\partial q}{\partial t^*} = g_i(q, q^*)\tau_s \\
\frac{\partial c(x^* = 1, \forall t)}{\partial x^*} = 0 \\
\frac{\partial c(x^* = 0, \forall t)}{\partial x^*} = Pe(c - c_{inlet}) \\
c(x^* \in (0,1), t^* = 0) = c_0 \\
q(x^* \in (0,1), t^* = 0) = q^*(c_0)\\
q^* = f_j(c, p)\\
\tau_s = L/v
\end{align*}

Where $g_i(q, q^*)$ and $f_j(c,p)$ can be:
\begin{align}
g_{LDF} = 0.22(q^* - q) \label{ap_1} \\
g_{\textrm{Vermeulen's}} = 0.22\frac{q^{*2} - q^2}{2.0q} \label{ap_2}\\
g_{\textrm{iLDF}} = 0.22(q^* + 0.2789q^*e^{{\frac{-q}{2q^*}}} - q) \label{ap_3}\\
f_{\textrm{Langmuir}} = \frac{55.54\times1.8c}{1.0 + 1.8c}\\
f_{\textrm{Sips}} = \frac{55.54\times1.8c^{1.5}}{1.0 + 1.8c^{1.5}}
\end{align}

in which, $t^* = t/ \tau_s$ is the dimensionless time, $x^* = x/L$ is the fixed bed dimensionless length, the axial distance from column inlet, $c$ is the liquid phase concentration, $q$ is the solid phase concentration, $q^*(c,p)$ is the solid phase equilibrium concentration (calculated from a known isotherm $f$ with known parameters $p$), $\varepsilon$ is the bed porosity, $v$ is the interstitial velocity, $Pe$ is the Peclet number.

\bibliographystyle{unsrt} 
 \clearpage
\bibliography{references}  

\end{document}